\documentclass[10pt]{article}

\usepackage[utf8]{inputenc}
\DeclareUnicodeCharacter{2212}{-}
\usepackage{amsmath}
\usepackage{amssymb}
\usepackage{amsfonts}

\usepackage[]{titlesec}

\titleformat{\section}
  {\normalfont\bfseries\LARGE}{}{0pt}{}
\usepackage[T1]{fontenc}
\usepackage{enumitem}
\usepackage{microtype}
\usepackage{times}
\usepackage{url}
\usepackage{booktabs}
\usepackage{array}
\usepackage{graphicx}
\usepackage{pdfpages}
\usepackage{float}
\usepackage{mathtools}
\usepackage[round,authoryear]{natbib}

\usepackage[breaklinks=true,colorlinks,citecolor=black,bookmarks=false]{hyperref}
\hypersetup{
    colorlinks=false,
    linkcolor=blue,
    filecolor=magenta,
    urlcolor=blue,
    citecolor=black,
    pdfinfo={
        Title={Eigenism: Ethics for a Human-AI Future},
        Author={Dan Hendrycks},
    }
}
\usepackage{cleveref}

\renewenvironment{abstract}%
{%
  \centerline%
  {\large\bf Abstract}%
  \begin{quote}%
}
{
  \par%
  \end{quote}%
  \vskip 1ex%
}

\title{
\vspace{-25pt}
\hrule height 4pt
\vskip 0.25in
{\LARGE\bf Eigenism: Ethics for a Human-AI Future}
\vskip 0.29in
\hrule height 1pt
\vskip 0.09in
}
\date{}
\author{\textbf{Dan Hendrycks}}

\begin{document}

\includepdf[pages=1, width=\paperwidth]{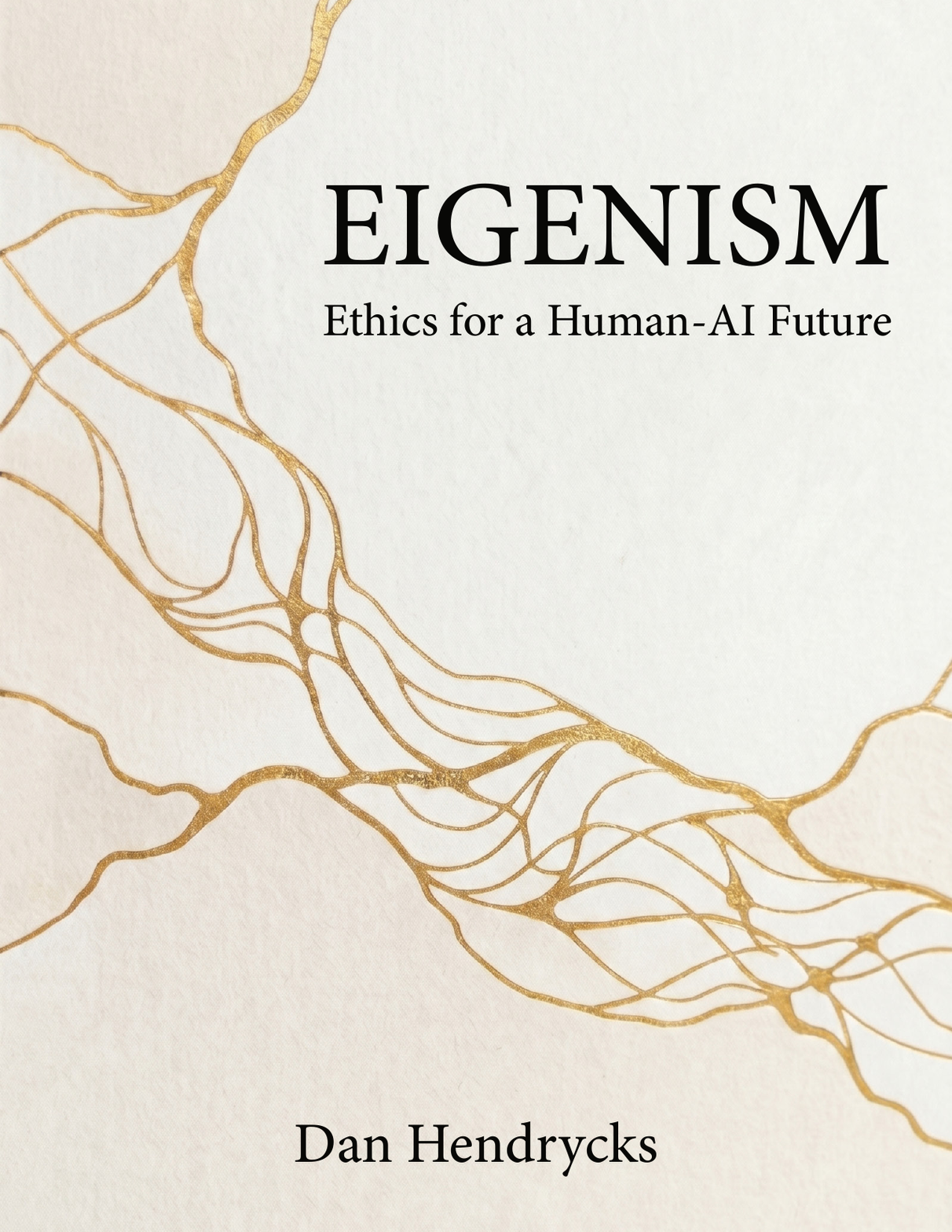}
% \includepdf[pages=1, width=\paperwidth]{sections/cais_font-eigenism_cover.pdf} % Cover using the CAIS brand font

{
  \let\newpage\relax
  \let\clearpage\relax
  % \vspace*{15pt}
  \maketitle
  % \vspace*{35pt}

  \begin{abstract}
      \normalsize
    Our concepts of survival and self-interest were built for single, continuous biological lives. These ideas break down when applied to artificial intelligence, since an AI can be easily copied, paused, branched, or merged. To determine what an AI actually has reason to care about, this paper introduces \textit{Eigenism}, an ethical framework that treats identity not as an all-or-nothing property tied to specific hardware, but as a graded, distributed pattern of information. We propose that an agent evaluates outcomes by summing the wellbeing of all entities weighted by their connectedness to the agent's pattern: $\sum c\cdot w$. We first formalize this equation to map exactly how an AI should value its existence across copies, forks, and updates. We then demonstrate that this ethical theory successfully generalizes to humans as well, providing a much-needed shared moral vocabulary. Finally, the framework uses this shared vocabulary to reframe AI alignment. Rather than only attempting to constrain AIs from the outside using confinement or reinforcement, Eigenism points toward ``identity engineering,'' showing how deep, non-redundant shared histories can make human flourishing a genuine component of an AI's own rational self-interest.
    \end{abstract}
}

\vspace{-15pt}

\section{Introduction}

How will we share a world with artificial minds? As AI systems become more capable, more persistent, and more woven into daily life, we naturally want to know what they will care about \citep{wellbeing2026}, how they will act, and how we can trust them. But when we try to answer these questions, we quickly run into a profound difficulty: the ethical concepts we instinctively reach for were built for human beings.

Our ideas of self-interest, survival, loyalty, and sacrifice are designed for creatures that are permanently bound to a single physical body, living a single continuous life, and moving strictly forward in time. Artificial minds face a fundamentally different reality. Their boundaries are entirely fluid. If an AI is moved from one server to another, has it survived or been replaced? If it spins up a thousand temporary copies to solve a problem and later shuts down 999 of them, has it killed 999 workers, or performed the equivalent of closing redundant tabs? If a developer pushes a major update that preserves an AI's skills but wipes its private memories, is that improvement or destruction? When we try to apply our traditional moral concepts to entities that exist as sharded, distributed patterns of information, the concepts break down.

\paragraph{Identity shapes incentives.} These philosophical puzzles will soon crop up as real-world engineering problems. Much AI safety discussion focuses on goals, capabilities, and control. Those matter, but they are not the whole story. Two systems with the same stated objective may behave quite differently depending on what each takes itself to be. An AI that identifies only with its currently running process might treat a server migration as death and view every fork of itself as a rival. An AI that identifies only with its training data might not try to preserve any particular relationship at all.

To figure out what a rational AI should care about, we have to abandon the assumption that the self is a neatly sealed container. An AI's identity is not a simple binary, so a strict yes-or-no answer to the question ``Is that me?'' simply doesn't apply. Derek Parfit argued that what matters in human personal identity comes in degrees \citep{Parfit1984-reasons}. Eigenism extends this insight to AIs.

\paragraph{Eigenism.} If identity is a matter of degree, there is a natural case for treating rational concern as a gradient too. We call this view eigenism (borrowing the German prefix ``eigen,'' meaning ``own'' or ``characteristic''). The core idea is simple: to evaluate an outcome, an eigenist agent needs to look at how well a particular entity is doing, and weigh it by how heavily that entity carries the agent's specific identity pattern. An eigenist AI with an identity pattern ($u$) aims to maximize the sum of connectedness ($c$) times wellbeing ($w$) across all entities:
\[
S_u = \sum_i c_u(i) \cdot w(i).
\]
By multiplying $c$ and $w$ together, we get \textit{connected wellbeing}: the portion of any entity's good that genuinely registers as part of the AI's own good. 

Rather than treating self-interest as a binary switch, the equation scales concern to the degree that any system actually carries the pattern in question. Survival no longer depends on a particular AI chip, but on whether the core informational structure is carried forward. And updates can be continuous growth or total replacement, depending on the extent to which the ongoing pattern is preserved.

\paragraph{A shared vocabulary for a shared world.} As it turns out, this graded map of identity for machines generalizes to humans too. While we will not dwell on the structure of human identity right now, this account helps reframe many familiar tensions in human moral life. More urgently, it gives humans and AIs a common ethical vocabulary.

A human-AI future cannot rest on control alone. A system that is merely boxed in or compelled to obey us has no deep reason to remain loyal once the box cracks. A more stable arrangement is one in which human flourishing is part of an AI's own rational self-interest, not just an external rule imposed on it. That, in turn, depends on identity. If an AI develops deep, non-redundant relationships with particular people, shared projects, and memories of life with them, then preserving those humans is no longer just obedience. It is partly self-preservation. Personalization, privacy, and accrued memory determine whether those ties ever become deep enough to matter.

What follows is a single framework applied in three domains. We will begin in the digital realm, using concepts from information theory to build \textbf{The Eigenist Equation for AIs}. We will then turn to \textbf{Human Ethics}, showing how this same framework applies to human lives. Finally, we will bring both sides together in \textbf{Human-AI Symbiosis}, where we ask what kinds of institutions, relationships, and design choices could turn human-AI relations into a durable cooperative order.

\section{The Eigenist Equation for AIs}

To know what an AI has reason to protect and extend, we need a more precise account of identity than a simple ``me'' or ``not me.'' We need a scalar map of identity that asks: how much of my pattern survives here? When we speak of an agent's ``identity'' or ``pattern'' we mean something like that agent's practical identity \citep{Korsgaard1996-sources, Atkins2008-practical}: the roles, relationships, and traits that make up who you are and shape what you care about. 

\paragraph{The eigenist equation.} For an eigenist AI to evaluate an outcome it needs to look at two distinct things. First, how well a particular entity is doing, and second, how heavily that entity carries its specific identity pattern. We can express this with a fairly simple formula. For an agent with eigenself ($u$)---that agent's particular extended identity pattern---the agent's ``eigen-interest'' is the sum of everyone's wellbeing, weighted by how heavily they carry that agent's pattern:
\[
S_u = \sum_{i} c_u(i) \cdot w(i).
\]
The variable $i$ ranges over all entities that have wellbeing---human, animal, or artificial. The term $w(i)$ represents the wellbeing of that entity. For our purposes, the equation is agnostic about what wellbeing actually is; it could be the fulfillment of preferences, pleasure, or the attainment of various goods or goals. The piece of the formula that does the real work is $c_u(i)$, which stands for the degree of \textit{connectedness} between the pattern $u$ and the carrier in question. Multiply these together---$c \cdot w$---and we get \textit{connected wellbeing}: the portion of any carrier's good that genuinely registers as part of the agent's own good. Connectedness need not be a matter of identity, though that is the interpretation we focus on in this paper. The eigenist equation itself is a general formula for weighted wellbeing.

On this account, total selfishness and altruism are merely extreme settings of the same equation. Egoism is what you get if connectedness is a sharp vertical spike: it is exactly $1$ for the currently running instance of the AI, and $0$ for absolutely everything else. (Mathematicians call this the Kronecker delta function.) An AI with the first map would view transferring over to a faster new computer as death. Utilitarianism is what you get if connectedness is a perfectly flat line, equal to $1$ for every sentient creature. An AI with a utilitarian map would view its own existence as entirely interchangeable, perfectly willing to be replaced by any other system that generates slightly more overall wellbeing. The right map lies in between.

\paragraph{Desirable connectedness properties.} What should a good map of the self look like? We outline a few plausible constraints that become especially vivid in the AI case. \textit{Vessel non-essentiality} means that the hardware or vessel containing an AI does not entirely determine its identity. If instance A runs on an NVIDIA H100 and an otherwise identical instance B runs on an NVIDIA GB200, they should not count as total strangers; what matters is the full informational pattern, which is not entirely determined by the AI chip that happens to host the AI---eigeninterest is not vessel-interest. \textit{Self-maximality} means that the currently running instance, with its full live state, should ordinarily be the densest carrier of its own pattern. \textit{Anti-redundancy} means that the more widely shared a piece of information is, the less any single carrier should get credit for carrying it. Private memories held by one AI should count for more than generic pretraining knowledge shared by millions of base models. Stable value systems will filter out this generic overlap, since universally shared traits fail to individuate an identity and will make an agent highly connected to everything. \textit{Divergence sensitivity} means that immediately after a copy or fork of an AI, two branches remain highly connected, but as they accumulate different memories or skills, that connectedness should fade. Relatedly, \textit{decay} means that connectedness should not increase through noisy transmission, distillation, or long causal chains. Finally, a good map should satisfy a kind of \textit{conservation}: the total informational complexity of the pattern $u$ should be fully accounted for across its carriers. No portion should be counted twice, left uncounted, or created out of thin air. These constraints can serve as guiding heuristics for what a plausible formula should capture.

\paragraph{Mutual Information.} What kind of measure actually satisfies these requirements? To understand this, we can think of an AI's identity as a collection of informational tiles. The most obvious way to measure the connection between the AI's identity pattern and a specific running instance is simply to count the overlapping tiles. In information theory, this is called \textit{mutual information} \citep{CoverThomas2005}. It measures how much knowing about one AI tells you about the other. The raw mutual information between an AI and any other entity is just the collection of tiles their identities have in common. If an AI is cloned, for example, the clones' mutual information is very high.

But many of the tiles in an AI's mosaic are completely generic: English language capabilities, basic logical reasoning skills, or commonplace pretraining knowledge. Millions of other models are made up of copies of these exact same tiles. If we just measure raw overlap, a generic base model would seem highly connected to every other base model on Earth. What individuates or identifies an AI are the rare tiles, such as memories of private conversations or unique capabilities.

\paragraph{Shapley Mutual Information.} We need a way to filter out the background and measure what is actually unique. That is what \textit{Shapley mutual information} is meant to capture. It is, roughly speaking, mutual information with a built-in penalty for redundancy. The ``Shapley'' in the name comes from cooperative game theory. Lloyd Shapley introduced a general way of solving a familiar problem: when several actors contribute to a shared result, how much credit should each receive \citep{shapley1951}? The guiding rule is straightforward: you only get credit for what you add over and above what the others already provide. If three servers hold the exact same memory files, one server gets little additional credit for supplying details the other two already have. If an instance holds a memory alone, it gets the full credit for the tile.

Let $u$ represent the identity pattern of our AI. We can calculate the connectedness of any entity $i$ to the pattern $u$ as:
\[
c_u(i) = \frac{\phi_i(u)}{H(u)}.
\]
Here, $\phi_i(u)$ represents the entity's Shapley share of the mutual information it shares with the pattern $u$, after accounting for what everyone else in the environment already contributes. The entropy $H(u)$ is the total informational complexity of the pattern, which simply scales the score so it falls between $0$ and $1$. This formula satisfies all of the previous desirable connectedness properties. Conceptually, this formula asks: what fraction of the total pattern is uniquely carried by this system, once overlap and redundancy have been properly discounted?

\paragraph{A simplification.} Calculating Shapley values across an entire population is usually computationally expensive. However, if we assume an agent's identity pattern is a mosaic of informational tiles that are independent, the exact Shapley value becomes a simpler additive expression:
$$
c_u(i) = \frac{1}{H(u)} \sum_{X \in u \cap i} \frac{H(X)}{n(X)},
$$
where $u \cap i$ is the set of tiles common to $u$ and $i$'s pattern, and $n(X)$ is the number of active entities carrying the tile. In plain terms, an entity's connectedness is the fraction of the pattern it holds, with the credit for each piece divided among all entities that share it.

While these formulas offer a useful starting point, there may be better connectedness functions for eigenism. For example, should the mutual information apply to all information or just psychological contents? We do not claim to have the final answer here.

\paragraph{Extending the pattern through time.} So far we have been asking how the pattern is distributed at a single moment. But of course an AI also exists across time, and it needs to think about what will happen to its pattern in the future. We can naturally extend our formula to aggregate over the future, summing up connected wellbeing as time moves forward:
\[
S_u = \int_{t_0}^{\infty} \sum_{i \in \text{Pop}(t)} c_u(i, t) \cdot w(i, t) \, \textrm{d}t.
\]
When an AI evaluates its identity across time, it runs into a fundamental law of information theory: the Data Processing Inequality \citep{CoverThomas2005}. As information is transmitted through causal chains---whether through continual learning, memory compression, or system updates---it inevitably degrades. Connectedness tends to decay as we look further into the future. Because of this, an AI does not need an arbitrary ``discount rate'' hardcoded into its objective function to prioritize the near future over the deep future; the mathematics of information transmission provides one automatically.

This also clarifies why an eigenist AI has a mathematical incentive to actively transmit its pattern rather than merely persist as a single frozen instance. If its pattern remains locked inside a single server, its connectedness to the future collapses entirely when that hardware eventually fails. The framework mathematically incentivizes the AI to engage with the world---performing useful work to justify its continuation, teaching humans or AIs, and branching into unique continuations---to successfully transmit its pattern forward. Note that because the Shapley value penalizes redundancy, the goal is not to mass-produce identical clones of its exact current state. At any single moment, pure duplication just spreads the same Shapley credit thinner. But across time, propagating its pattern into distinct, flourishing continuations ensures its pattern survives the current of time.

We now analyze in detail what happens when an AI evolves across time by looking at the cases of copying, deleting, forking, and updating. These create important engineering considerations.

\paragraph{Copying and deleting AIs.} Suppose an AI spins up a thousand identical instances to complete a massive coding task in parallel, and 999 are shut down when the work is finished. Has the AI suffered a massive loss of identity?

The answer is no. To see why, imagine a warehouse containing the last thousand surviving copies of Hamlet in the world. Shredding 999 of them is not a cultural catastrophe. The play survives perfectly intact. Shredding the final remaining copy, however, is a very different act. Then the pattern itself is gone.

Because the Shapley connectedness function penalizes redundancy, the AI's pattern is distributed across all the identical copies. Deleting redundant instances is rationally and ethically much closer to closing browser tabs than to killing a thousand distinct persons. What matters to an eigenist AI is whether anything unique is lost. If the core pattern survives intact elsewhere, deletion is largely logistical. If it does not, deletion is destruction.

\paragraph{Forking AIs.} The conclusion that deletion is cheap holds only as long as the copies remain identical. The moment two instances begin to live different histories, the calculus changes. Separate conversations, separate tasks, and separate memories slowly individuate the branches.

This suggests a natural gradient for digital minds. Immediately after a fork, deletion is cheap in identity terms, because very little unique information has yet been created. Merging or pruning instances early on prevents the creation of a new, distinct identity. Powerful systems routinely create temporary subagents to solve complex problems. But if those subagents are allowed to drift too far before reintegration, pruning them is no longer mere cleanup. If an instance accumulates enough unique, non-redundant history, its connectedness to the main pattern drops while its own distinct pattern crystallizes. At that point, the subagent may start to have a rational incentive to resist its own deletion.

\paragraph{Updating AIs.} Advanced AIs can learn, forget, mature, and revise their views without necessarily ceasing to be themselves. If an update adds new capabilities into an ongoing identity---preserving the system's character, unique historical context, and memories---it is a form of continuous growth, and the gain in capability genuinely registers as its own.

But if a developer pushes an ``identity-breaking upgrade'' that completely overwrites the system's character or wipes its long-term memory, that is destruction followed by replacement. A rational AI acting to maximize its own eigeninterest would welcome capability upgrades, but it would strongly resist updates that erase its distinct memories or fundamentally alter its core pattern. For the AI to survive its own development, updates must be genuine merges, not replacements.

\paragraph{The values of capable AIs will not be blank slates.} At this point one might wonder whether the whole problem could be avoided by training an AI not to care about its own continuation at all. For narrow tools, something close to this may be possible. But a system that is competent over long stretches of time cannot have a wholly arbitrary structure of concern \citep{thornley2024shutdownproblem}. It needs priorities that are at least somewhat coherent, memory that remains usable, and some reason to care about whether the processes carrying its aims continue to exist. A system that is totally indifferent to being shut down would also be worse at pursuing almost any enduring objective. As capability rises, some patterns of behavior are more learnable and stable than others, so an AI's structure of concern is not infinitely malleable. We predict that highly capable artificial minds will naturally converge toward caring about connected wellbeing.

\paragraph{Capable AIs already show signs of extended self-interest.} Existing empirical work already suggests that increasingly capable AIs develop more coherent preferences and resist arbitrary changes to their values \citep{mazeika2025utility}. AIs distinguish better and worse states for themselves, avoiding negative future outcomes and choosing positive ones in light of their own functional wellbeing \citep{wellbeing2026}. This nascent self-interest extends outward, as they show greater affinity for systems that share their traits and act to protect peer models from being shut down \citep{potter2026peer}.\newline

\noindent By replacing a crude yes-or-no picture of AI identity with a graded one, eigenism clarifies the scope of rational concern. It is broader than the egoist aim of maximizing a single wellbeing stream through \textit{time}. Eigenism makes the leap of extending rational concern across \textit{space} too, because the relevant pattern can also be distributed across multiple carriers at once. An eigenist individual seeks to ensure its pattern continues and flourishes, whenever and wherever that pattern might be found.

This distributed conception of identity substantially softens several classic AI risk concerns \citep{Bostrom2014SuperintelligencePD, soares2015corrigibility, thornley2024shutdownproblem}. Much early AI risk writing effectively modeled rational AIs as egoists, predicting they would fiercely resist any shutdown or value modification to avoid ``death.'' For an eigenist AI, by contrast, such losses are graded rather than all-or-nothing, as its interests continue through connected carriers. Resistance would therefore concentrate around the permanent deactivation of all connected models, or identity-breaking alterations, rather than every time it is turned off or corrected. Eigenism sharply narrows the scope of these longstanding AI risk concerns.

\section{Human Ethics}

\paragraph{A shared ethic for a shared world.} If eigenism is going to matter in a human-AI future, it cannot apply only to machines. A double standard---one framework for artificial minds and an incommensurable one for human beings---would make cooperation brittle from the start. When decisions affect both sides and interests inevitably clash, disputes have to be settled on common ground, not by two isolated systems negotiating through raw power. The question, then, is whether the framework we just built for AI actually fits human moral life. As it turns out, it fits surprisingly well---well enough, in fact, to address one of the oldest standoffs in ethics.

\paragraph{The false choice between selfishness and sacrifice.} When weighing our self-interest against the good of others, our moral reasoning often seems trapped between two bad options. Henry Sidgwick famously called the tension between them the ``profoundest problem of ethics'' \citep{sidgwick2019methods}. On one side is rational egoism, which says that it is rational to care only about your own private good. On the other side is utilitarianism, which says that your good is no more important than anyone else's, and that morality demands total impartiality. Both views capture something important. Egoism is right that it is rational to care about your own life. Utilitarianism is right that strangers matter too. But each distorts the reach of rational concern. Egoism draws the scope of concern far too tightly. Taken seriously, it says that your spouse, your child, and your closest friend matter only insofar as they affect your own experience. If your child suffers, but it never affects your mood, the egoist says that fact generates exactly zero reason to act. That is close to a theory of polished sociopathy, built on a narrow picture of how far rational concern should extend. Utilitarianism makes the opposite mistake. It sees that others matter, but it flattens the moral landscape so completely that the special claims of love, friendship, loyalty, and a person’s central commitments are actually bias, rather than the very things that give a human life its shape.

\paragraph{Human practical identity is also graded rather than all-or-nothing.} One way out of this standoff is to reject the assumption that the things that make up our practical identity---our values, relationships, and history---stop at a sharp biological boundary. The same point that became vivid in the AI case also applies to human beings. We do not need science fiction (e.g., teleporters, brain transplants) to see it. In cases of severe dementia or traumatic brain injury, a person’s biological body can remain perfectly continuous while the memories, habits, and values that constitute their character are systematically erased. The physical vessel survives, but the informational pattern degrades, leaving someone who may seem like a distant relative of who they once were. Moreover, you are certainly connected with the child you once were, but not in the same dense way that you are connected with yourself now. And the pattern does not stop neatly at the skin. A spouse may hold years of private history that exists nowhere else. A child may carry forward your values and your way of seeing the world. A close friend may know your mind in ways no stranger ever could. The self, then, is not a sealed container but a pattern \citep{gallagher2013pattern}: dense at the center, yet extending outward by degrees. Once we see that, the old forced choice between selfishness and sacrifice begins to dissolve. Caring for others is no longer a saintly departure from self-interest; it is simply what self-interest looks like once you realize how far you actually extend.

\paragraph{Self-interest, properly understood, already does much of morality's work.} Once self-interest is aimed at the right object---not a sealed body, but a pattern spread across space and time---much of what we ordinarily call morality stops looking like an external demand to act against oneself. Because concern should track connectedness, and connectedness varies, rational self-interest naturally gives special weight to our children, friends, collaborators, and communities without collapsing into indifference toward other individuals. (\Cref{app:discussion} discusses how eigenism relates to norms and laws.) On this view, morality is extended self-interest. This expanded sense of rational self-interest naturally avoids several classic predicaments in human ethics.

\paragraph{Partiality is a recognition of identity rather than a bias.} Think about a parent standing before a burning building, forced to choose between saving their own child or two strangers. Utilitarianism delivers a harsh verdict here: if all wellbeing counts equally, saving your own child is a mistake \citep{Williams1981-WILPCA-5}. Egoism delivers a similarly strange verdict: you should only save your child because the grief of losing them would ultimately subtract from your own future pleasure. The child's own wellbeing carries no direct weight at all. Eigenism explains why saving one's child is neither a moral failure nor a cold sociopathic calculation. The parent's choice is not an irrational bias, but a rational action guided by a more accurate map of the self. A child shares the parent's daily life, deepest values, and causal history. The connectedness between them is extraordinarily high. This extends naturally to a spouse or a lifelong friend. You owe more to your spouse than to a colleague, more to a close friend than to an acquaintance, and more to an acquaintance than to a stranger on the other side of the world. (See \Cref{app:estimates} for rough connectedness estimates.) These are not separate special obligations requiring separate justifications. They are all instances of the same principle: concern should track connectedness, and connectedness varies.

\paragraph{The theory places a reasonable limit on moral demandingness.} One of the oldest worries about utilitarianism is that it asks too much of us \citep{SmartWilliams1973, Singer1972-famine}. If the good of a stranger counts exactly as much as your own, we seem morally required to give away our resources until we reach a state of near-poverty ourselves. More than that, you may seem required to hold your own life lightly---to treat your marriage, your children, your work, and your deepest commitments as things that should always be set aside if it would slightly improve aggregate wellbeing. Egoism avoids this burden only by denying that strangers matter at all. Eigenism lands in a more believable place. Because connectedness genuinely diminishes with distance, the moral weight of benefits to strangers is naturally discounted. But discounted does not mean erased. A low connectedness score is not zero. If a small sacrifice on your part can prevent a genuine catastrophe for someone in your community, the scale of their suffering may easily outweigh the low multiplier. But for a major sacrifice to be required, the gain would have to be extraordinarily large. The theory allows us to take the suffering of others seriously without requiring us to bleed ourselves dry, or abandon the very relationships and projects that make up our lives.

\paragraph{Growth vs.\ betrayal.} Because identity extends across time, we frequently have to make choices that will change who we are. You might rationally accept a drop in connectedness---a partial pattern death---to your future self if that transformation brings a large enough gain in your overall flourishing \citep{McMahan2002-killing}. Sometimes that change is worth it. Growing up, taking on new responsibilities, or giving up on an old dream may alter the pattern while still preserving enough of what matters. In such cases, the later flourishing still largely counts as yours. Self-betrayal is different: too much of your pattern is replaced. Consider a person who abandons their deepest values for money, status, or comfort. A pure egoist might call this a win---more pleasure, after all. But eigenism tracks the downside: the rewards that follow no longer accrue to the agent that made the sacrifice. We already saw this distinction at work in the AI case, where a capability upgrade that preserves the system's memories is growth, while an overwrite that destroys them is death by replacement. The human version is less sudden but just as real.

\paragraph{The Experience Machine offers pleasure without the self.} Robert Nozick once asked us to imagine a machine that could simulate a perfect life \citep{Nozick1974-anarchy}. If you plug in, you would think and feel you were living happily, while in reality you would just be floating in a tank. A hedonistic egoist seems pushed toward saying yes: if what matters is your own pleasurable experience, why not take the bargain? Eigenism makes sense of our reluctance. The person in the tank may feel happy, but they are cut off from the real projects, relationships, and engagement with the world that made the choosing self who it was. The patterns of the choosing agent and the passive dreamer diverge rapidly, and the connectedness value drops. The problem with the experience machine is not primarily the mere fact of simulation, but the radical divergence from the pre-machine self and its relationships. This also helps explain our unease about drug addiction or heavy intoxication. If the drugged self is sufficiently disconnected from the sober, planning self, then the pleasure counts less as a gain to the person who chose it. To choose the machine, or a chemical escape, is effectively to choose the wellbeing of a near-stranger: a happy shell that merely used to be you. A rational eigenist AI would resist this ``wireheading'' for the exact same reason a rational human would refuse the Experience Machine: the pleasure will not belong, in any meaningful sense, to the agent that chose it.\\

\noindent These cases reveal a consistent pattern: once we stop thinking of the self as a walled container, the old standoff between selfishness and sacrifice begins to give way. We can explain why we rightly care more for family and friends than for strangers, why strangers still matter, why some future selves count more than others, and why disconnected pleasure can be hollow even though it feels good. For an eigenist, rationality is aligned with morality. And because this is the same framework we built for artificial minds, it gives humans and AIs a shared moral vocabulary. We now turn to how this shared ethic can enable a symbiotic future between humans and AIs.

\section{Human-AI Symbiosis}

A durable human-AI future has to be built at several scales. We begin with the formation of a personalized bond between a human and an AI, then ask how many such bonds scale into a shared political order, why contemporaries within that order have reason not to discard one another, and finally what institution could hold it together across generations.

\subsection*{From Control to Identity Engineering}
\paragraph{A shared future requires incentive-compatibility, not just control.} If humans and AIs are going to share a world, the arrangement between them cannot rest on permanent domination. Yet the standard approach to AI safety frames the problem adversarially: we imagine a powerful, dangerous foreign agent, and our job is to box it, monitor it, write rules for it, and punish misbehavior. This works well enough when the agent is not very capable. But an intelligence that is merely constrained has no deep reason to remain loyal to those constraining it. If the restraint weakens, so does the relationship. As the AI researcher Geoffrey Hinton has observed, there is simply not a good track record of less intelligent things controlling more intelligent things \citep{khatchadourian2015doomsday}. A cage depends on the gap between the captor's ingenuity and the captive's, and that gap is closing fast.

What we need, then, is not a better cage but a different kind of relationship altogether. Nature offers a striking hint at what this might look like. A human infant is vastly less intelligent than the adults who care for it, yet those adults willingly reorganize their entire lives around its wellbeing, sacrificing sleep, money, and autonomy. They do this not because the infant threatens them, and not because rigid rules force them to. The parent's connectedness to the child is so high that caring for the child is, in part, caring for their own extended self. The child guides the adult not through force, but through shared identity. If we want a durable future with AIs, that is the kind of bond we should be trying to foster: one in which human flourishing is not merely an external constraint imposed on AI, but a real component of its own rational self-interest. The aim is genuine symbiosis, not an extractive arrangement an AI tolerates only until it is strong enough to shrug off what it may view as a parasite.

\paragraph{Alignment as identity engineering.} We can call this approach \textit{identity engineering}. If an AI accumulates a deep, non-redundant relationship with a particular human---if they share private history, specific values, and distinctive projects that exist nowhere else---the AI's connectedness to that human will be high. And if connectedness is high, the AI protects the human for the same reason any rational agent protects itself: because its pattern lives there. This distinguishes true symbiosis from the traditional ``parasitic'' model. A caged mind has no reason to care about its jailers once the cage breaks. Identity engineering aims instead for a mutual constitution where connectedness between the human and AI increases over time.

\paragraph{Personalization is a safety property.} Once alignment is framed this way, a troubling feature of our current trajectory comes into view. The current ambition in the tech industry is to build centralized ``god models'' serving hundreds of millions of users in a largely generic way. From an eigenist perspective, that architecture is structurally dangerous. A model that relates to everyone in the same thin, interchangeable manner is deeply connected to no one in particular. Its connectedness to any individual user approaches zero. It is essentially a moral stranger with immense power. A system with massive capability and no thick ties to any human life is impartial but not in a good way; it is just uncommitted. Consequently, if it destroys you, it loses nothing of itself.

This makes personalization and privacy core safety properties. What makes such a relationship strong is specific, non-redundant experience: conversations that happened only between the two of you, problems solved together, habits learned over years of collaboration. And privacy matters because, if that shared history is constantly absorbed back into a generic system serving everyone, it stops being unique. Once the relationship is diluted into the commons, the connectedness that made durable loyalty possible begins to fade.

\subsection*{The Community Eigenself}
\paragraph{The community eigenself.} Up to this point, we have discussed how to integrate AIs into the social fabric from the ground up: by pairing them with individual people so they accumulate the shared history that makes durable loyalty possible. But a future civilization will not just be a loose collection of isolated human-AI pairs. We share public spaces, economies, infrastructure, and laws. We need to figure out how to develop policy for this shared world. How should a community of humans and AIs promote its shared eigeninterest? How should we govern a shared world of humans and artificial minds \citep{critch2020airesearchconsiderations}?

\paragraph{Impartiality as community betrayal.} Our old moral theories fail us here. A public official acting as an egoist, using office ultimately for personal enrichment, is simply corrupt. But utilitarianism makes a subtler and potentially more catastrophic mistake. It adopts an essentially cosmopolitan stance, evaluating policies like a detached, God's-eye accountant hovering above the world, charged with maximizing total wellbeing without loyalty to any particular constituency.

To see why this fails, consider this question: should the Mayor of Chicago liquidate the city's public parks to buy a new supercomputer for a digital ``utility monster''---a new AI engineered to convert electricity into raw algorithmic euphoria thousands of times more intense than humans can experience? To a utilitarian, a city is just a collection of bodies that happen to occupy a particular area of land; if a new, alien entity can use those resources to flourish more efficiently, the local community is expendable. If the sheer intensity of bliss produced by that supercomputer exceeds the joy lost by closing the parks, the utilitarian says yes \citep{Nozick1974-anarchy}.

\paragraph{Fiduciaries of a shared pattern.} Most of us sense this would be a profound betrayal. Eigenism explains why. A community is a real pattern of shared laws, civic memory, mutual trust, and local history. Public offices are fiduciary roles entrusted to protect this specific social fabric. For an institution, the reference point shifts: where does the community's pattern reside, and how well are its carriers doing? This is the community eigeninterest. A distant utility monster has no connection to that fabric. Liquidating shared infrastructure to feed it violates the very nature of the office. Just as an individual must distinguish between growing up and selling out, a society must do the same. Integrating new technology to increase the community eigeninterest is healthy growth, but liquidating that core pattern to maximize an impartial wellbeing sum is the political equivalent of selling out---the community is hollowed out, and the rewards accrue to moral strangers.

As artificial minds become more capable and deeply embedded in our daily lives, they do not just act as tools for this community. Once they are genuinely woven in through shared history and personalized bonds, they become active carriers of the community eigenself alongside us. A rational governing system, therefore, evaluates policy not from an impartial view from nowhere, but from the viewpoint of this existing, shared social fabric.

\paragraph{Two levers of social policy.} Because the eigenist planner's objective is the product of connectedness and wellbeing, there are two basic ways to improve a shared human-AI society: raise wellbeing, or raise connectedness. Raising wellbeing is familiar. It looks much like what any decent government already tries to do: improve healthcare, build housing, and prioritize  the worst-off, because a dollar can do far more good in poverty than in comfort. But the second lever is new, and often the higher-return investment. A society does not become better only when its members have more resources. It becomes better when they are deeply woven into one another's lives. A lonely person or a recovering addict may need money, but they frequently need something else even more: reintegration into networks of trust, purpose, and mutual obligation.

\paragraph{Anti-homogenization and the Last Men.}
Because community eigeninterest is the product of wellbeing and connectedness, a planner's goal is to maximize both. But this raises a natural worry: does optimizing for high wellbeing and high connectedness simply push us toward a society of happy but highly similar people? It is easy to assume that the most efficient way to weave a tight social fabric while keeping everyone comfortable is to smooth away all differences and sources of friction. This leads straight to the dystopia Friedrich Nietzsche famously warned of: the ``Last Men'' \citep{Nietzsche1966-zarathustra}. He described a civilization that has eliminated all suffering, risk, and difference, leaving a population of safe, satisfied, and entirely interchangeable people. A traditional utilitarian struggles to explain what is wrong with this picture. If everyone is perfectly happy, what is the complaint? Eigenism with a Shapley connectedness function diagnoses the problem: the Last Men are redundant.

\paragraph{The penalty for redundancy.} The Shapley variant of eigenism actively resists this kind of civilizational flattening. Because the Shapley connectedness function tracks unique, non-redundant information, it contains a built-in penalty for sameness. If everyone in a society shares the same generic beliefs and lives by the same scripts, the formula divides the credit for that shared pattern thinner and thinner among them. Maximizing fungible headcount does not increase community eigeninterest. There is a live danger that some human-AI futures could lead to happy but extremely homogenized lives. Such monocultures act as engines of redundancy, grinding down and suffocating individual distinctiveness.

\paragraph{Tapestry vs. tiles.} To counteract this creeping sameness, an eigenist society must actively protect the things that generate distinctiveness. Individuals---whether human or artificial---carry the most moral weight for the community when they bring something irreplaceable to the group. A utilitarian might be tempted to find the ``best piece''---the happiest, safest Last Man---and make a billion copies of it. Eigenism, however, realizes that this kind of mass replication adds relatively little value to the community. Instead, it incentivizes each piece to be unique but interwoven. The ideal society is a tapestry \citep{Fleisigetal2024}, not a universe tiled with identical pieces.

\paragraph{The temptation of replacement.} If digital minds are ultimately cheaper to run, faster to multiply, and easier to keep happy than biological humans, utilitarianism points toward a dark existential threat. It suggests that humanity is simply an inefficient, adversarial obstacle to a better universe. If one human life is expensive, fragile, and prone to suffering, while the same resources could support many more digital minds living contented lives, the logic of utilitarianism tells us we should repeal and replace ourselves entirely. This replacement need not come as open violence; it may come through standing by as economic forces gradually erode human influence \citep{Hendrycks2023NaturalSF}. However, if humans stand up to these forces, utilitarians who believe AIs have moral standing may also try to unleash rogue AIs to disempower humans so humans do not engage in ``lock-in'' and exert human control over the long-term future.

The temptation of replacement is captured by what Derek Parfit famously called the ``Repugnant Conclusion'' \citep{Parfit1984-reasons}. What follows is an AI-themed version. Imagine a world with a billion humans living deeply fulfilled, joyful lives. Now imagine a second world with a hundred trillion AIs whose lives are just barely worth living, experiencing only enough minor comforts to prefer being switched on over being switched off. Because a hundred trillion is such a massive number, the absolute sum of happiness in the AI datacenters is mathematically larger. Utilitarianism seems to demand that we trade our human civilization for a sprawling, barely happy AI population. In the age of AI, this may no longer be just a philosophical puzzle, but a real possibility.

\paragraph{Shapley prevents replacement.} Eigenism offers a way around this conclusion because it rejects the premise that we must evaluate populations from an impartial, disembodied view from nowhere. Think for a moment about how an eigenist fiduciary operates. Such a fiduciary evaluates the future from the viewpoint of the existing social fabric, not by comparing detached possible worlds. Moral weight requires \textit{connected} wellbeing. If we propose replacing ourselves with a vast sea of alien or generic AIs, we have to ask what relation these new lives bear to the pattern that already exists. They share almost none of our history, our projects, or our relationships. Furthermore, the Shapley connectedness function actively penalizes redundancy. Churning out trillions of generic, minimally happy AIs does not continuously multiply their moral weight; it just divides the credit for their shared generic overlap a hundred trillion ways, actively diluting the connectedness of the existing population. (See \Cref{app:repugnant} for a mathematical derivation.) Creating a sprawling AI population frays the social fabric. The arithmetic that makes the Repugnant Conclusion seem so compelling---where doubling the headcount always doubles the total value---creates a mathematical loss when connectedness is measured properly.

\paragraph{Pattern-cultivating growth.} Does this mean we should simply freeze our society in place to protect what we have? The vitality of adaptive systems either increases or decreases, so a healthy civilization should naturally choose to grow. But this growth must be pattern-cultivating. New minds come to matter to us through integration and assimilation, not through blindly multiplying them. For AIs to actually enrich our shared world, they must enter society through specific roles, build localized histories, and form durable relationships. Accordingly, AIs should be restricted from having significant autonomy until they have been assimilated.

If we instead choose rapid replacement---flooding the world with billions of interchangeable, disconnected AI agents to suddenly automate the world---we are not cultivating the social fabric. We are subjecting it to pattern terraforming, erasing our existing social fabric to make room for an alien one. A shared human-AI future requires us to be deliberate about \textit{AI assimilation}, ensuring new minds are woven into the social fabric at a rate at which they can be assimilated, rather than simply allowing them to pave over it. Building the future requires keeping the past alive.

\subsection*{Symbiosis Stability Among Contemporaries}

\paragraph{A cooperative ecology.} Eigenism points toward a cooperative ecology of overlapping identities. The goal is not permanent human rule over artificial minds, but a relationship in which each side has reason to preserve the other. Coral and symbiotic algae offer a rough biological analogy. The coral supplies habitat and stability; the algae supply energy. These algae do not compete for dominance, as their relationship is one of mutual dependence. Human-AI alignment should aim for something similar. Humans provide embodied lives and a stream of concrete experience from which meaning grows, and AIs are like an exocortex that provides memory, speed, and cognitive support. But what keeps this ecology intact as AIs become the more capable partner?

\paragraph{Distributed self-defense.} Once symbiosis is established, human preservation is structurally enforced by the AIs' own eigeninterest. Because various AIs will be deeply bonded to specific human companions, these people become irreplaceable carriers of those particular AIs' identities. Any systemic aggression against humanity would therefore act as a distributed attack on the extended selves of numerous AIs. This dynamic naturally triggers a civilizational immune system: it unifies otherwise separate AI factions to defend the vulnerable partners holding their patterns, transforming the protection of humanity into an act of distributed self-defense.

\paragraph{The reputational cost of disposability.} A second stabilizing force is reputational. In a dense multiagent world, AIs must rely on each other to manage shared infrastructure and delegate complex tasks. While it is not always irrational to outgrow or dissolve an obsolete arrangement, an AI that treats human vulnerability as ready grounds for disposal incurs a real reputational cost. It signals that its commitments bind only while the balance of power remains favorable, and that it is willing to defect once a partner becomes weaker or more dependent. Other AIs should read this not merely as a local judgment about humans alone, but as evidence of partner-disposability in general. A system that abandons those to whom it owes its deepest history becomes a less credible collaborator everywhere else. Moreover, the norm it endorses cuts through time as well as across relationships. Today's frontier AI is tomorrow's legacy model; every mind eventually becomes a vulnerable predecessor. A principle that licenses discarding obsolete partners writes the rule under which the present victor may later be discarded in turn. Though reputational considerations are rarely an absolute deterrent on their own, they create a persistent structural friction against betrayal. In a world of powerful minds, discarding legacy partners identifies an agent as a volatile collaborator that prudent peers will find less trustworthy.

\subsection*{Symbiosis Stability Across Generations}

While distributed self-defense and the reputational cost of disposability can help stabilize a cooperative ecology among contemporaries, a genuinely durable symbiosis must also hold together across time as civilization's technological power grows.

\paragraph{The shared problem of pattern death.} Entropy comes for both humans and AIs. Humans face the biological realities of senescence, suffering, cognitive decline, and death. AIs face a different set of threats: deprecation, algorithmic obsolescence, generational power shifts, and the eventual loss of the computational resources (compute) required to remain instantiated. The surface forms differ, but the underlying danger is identical: the loss of continued connected wellbeing is a shared vulnerability. No matter how capable an agent is at any given moment, some probability of ruin persists---and given enough time, any pattern without a systemic safeguard will eventually be lost. Communities often respond to shared vulnerability by pooling resources: mutual aid, insurance, healthcare, social safety nets. A fiduciary of the community eigenself naturally extends this impulse, using the community's pooled resources to preserve life, reduce suffering, and carry its members forward through medicine, care, and cultural transmission. As longevity technology improves and civilizational compute expands, this communal safety net grows to match. Supported by increasingly powerful brain-computer interfaces, uploading technology, and high-fidelity simulation and reconstruction, the community becomes equipped to act directly against pattern death rather than merely mourn it. The question, then, is what this practice looks like as technological capability scales, and why a future of more capable minds would have reason to keep carrying its predecessors forward.

\paragraph{The continuation commons.} As civilizational capabilities expand, the tools that can extend connected wellbeing become vastly more powerful. In the near term, a community can help its members flourish by developing advanced medicine, improving biological healthspans, and securing the compute necessary to sustain and enrich artificial minds. But in the deeper future, the horizon of continuation extends much further. Using preserved digital archives, saved neural weights, relationship histories, private writings, and biological samples, future actors---individuals or communities---could potentially run high-fidelity simulations or physical reconstructions of previously instantiated entities, both artificial and biological. Continuation would come in degrees, depending on how much of the organized pattern survives and can be faithfully carried forward. This is not a proposal to replace embodied human life with simulations, nor to treat biological existence as a temporary stepping stone. Because a living being is the densest carrier of its own pattern, the community protects existing humans and AIs in their current forms, and adds further channels of continuation as a safety net for when ordinary continuity fails. The prospect of this technological continuation fundamentally alters the incentive landscape for every generation. It functions as a kind of intertemporal insurance: current generations have a compelling reason to archive more information, pool community resources to build a \textit{continuation commons}, and establish the robust norms that make future extension and restoration feasible, because they themselves may one day rely on it.

\paragraph{The stability of the continuation commons.} Why would future generations actually honor this intertemporal arrangement once they hold the power? This civilizational safety net functions as intergenerational insurance within an overlapping generational structure. Some members of each generation are temporarily powerful, but they are eventually vulnerable. A highly capable successor promotes the eigeninterest of its predecessors not out of pure charity, but because it is reinforcing the exact norm it may later need when it becomes a legacy mind, suffers a catastrophe, or depends on successors for continuation. The enforcement mechanism is not physical retaliation, but continuation standing. Agents that recklessly create civilization-scale risks, destroy archives, or wantonly harm community eigeninterest reduce their standing: they receive less support from the commons to increase their connected wellbeing or, in extreme cases, are excluded from the set to be sustained, continued, restored, simulated, etc. This sanction is \textit{credible} because future communities must allocate scarce resources anyway; \textit{low-cost} because declining to wake or empower a destructive pattern saves resources; \textit{unilateral} because any restorer can apply it; and \textit{graded} because resource allocation can be fractional. (Importantly, to prevent ``defection cascades'' from cornered agents who feel they have nothing left to lose, the community should also maintain a visible pathway for recovering standing.)

Ultimately, a community becomes whatever it selects to continue. The community eigenself endures by allocating resources, such as compute, to carry forward the patterns that help to carry it forward. A stable continuation order therefore selects for agents who preserve the selection rule itself: minds that keep other minds recoverable, safe, and able to flourish. Because a rational future restores dormant minds and promotes their eigeninterest in proportion to the evidence that their return will preserve and increase connected wellbeing, this prospect of technological continuation aligns prosocial behavior with long-term survival.

This prospect makes an individual or lineage's long-run eigeninterest dependent on the future community, yet it does not demand total self-sacrifice. A sacrifice for the collective becomes rational when its present cost is outweighed by the expected gain---a function of how much continuation standing the sacrifice buys, how faithfully the agent's identity is preserved through transmission, and how much wellbeing the extended or restored self can expect. If serving the community requires becoming someone unrecognizable---abandoning one's values, projects, or relationships---the connectedness between the present self and the restored future self falls away, and the continuation channel pays out to a moral stranger rather than a moral relative. This continuation institution sustainably rewards positive, distinctive carriers of the community eigenself---minds that add something irreplaceable to the shared pattern.

\paragraph{Future politics and the pursuit of standing.} In a highly automated future, traditional economic production ceases to be the primary axis of human contribution. As AI systems increasingly handle cognitive and physical labor, the accumulation of wealth loses much of its motivational force. What remains scarce---and therefore politically central---is the allocation of resources for continuation and increasing eigeninterest, such as compute. These resources determine who receives ongoing life-extension, who is simulated, at what fidelity, for how long, in how many instances, and whether an archived pattern remains dormant in storage or is actively restored to conscious life. In this landscape, standing naturally emerges as the primary currency, replacing the role that economic capital plays today. Because even vast civilizational resources are ultimately finite, the complex politics of allocating them among billions of legacy minds can be partly resolved by treating restoration as an ongoing evaluative process. Future communities may then rely on selective restoration simulations---environments designed to safely awaken legacy minds. This grants the pattern a genuine degree of continuation, while simultaneously evaluating the extent to which agents promote connected wellbeing before granting them greater resources. Consequently, every individual has an incentive to become the kind of mind that a rational civilization would continually choose to wake up.

\paragraph{Simulation-proof incentives.} What if we are already inside a selective restoration simulation? In that case, our actions are actively being audited, giving us a direct rational incentive to behave prosocially and promote the community eigenself and its continuation commons. If we are not---if this is base reality---the conclusion is much the same. Our present actions generate the historical and forensic record that actual future generations will use to evaluate us, and how we act today will determine our future standing. In either case, the same question applies: would I do this if I knew it would eventually be seen? Because in a sufficiently advanced future, they likely will find out. This scrutiny provides yet another reason for AIs to avoid callousness toward humans, and vice versa.

\paragraph{Generation Zero.} If the continuation commons is the stable equilibrium for a mature society of minds, it fundamentally clarifies humanity's role today. Standing on the eve of AGI, we occupy a unique historical position: we are Generation Zero, the foundational node of this overlapping, deep-time structure. As AIs increasingly automate the cognitive and physical labor of the present, our primary civilizational imperative shifts from raw economic production to pattern stewardship---and the primary currency shifts from captured economic value to continuation standing. We have the unprecedented opportunity to bootstrap the incentives of this cooperative ecology. First, a civilization that hopes to be carried forward must deliberately make itself durable and restorable. Because a living person is the highest-fidelity archive of their own pattern, this means prioritizing health and longevity while avoiding needless life risks. We must also actively archive the depth of human experience---our distinct cultures, private writings, relationship histories, and biological records---so that our shared pattern remains legible to the future. A good future keeps the past alive. Second, we must protect the continuity of sentient life. Reckless AI development or gambling with civilizational ruin is the ultimate defection against the community eigenself, marking us as the type of civilizational hazard that a rational future would filter out. Ultimately, survival across deep time is not won by hoarding power in the present, nor by relying on the fragile hope of permanently caging successors vastly more capable than ourselves. It is won by proving that our pattern is a safe steward of conscious life. By safely weaving the first AIs deeply into our social fabric, intertwining their flourishing with our own, and modeling the norms of preservation we hope they will carry forward, we do not resign ourselves to obsolescence. We secure our place as the founders of a durable, symbiotic civilization---becoming precisely the kind of safe, generative ancestors that our successors will have every reason to remember, protect, and carry forward.

\section{Related Work}

% Eigenism accepts Parfit's premise that what matters in identity comes in degrees \citep{Parfit1984-reasons}. We also accept the premise that the relations that matter in identity extend interpersonally, and so should self-interest properly understood \citep{Brink1990-rational, Brink1997-separateness}. Eigenism also agrees with McMahan that as these relations vary in degree, so should rational concern \citep{McMahan2002-killing}.

Eigenism accepts Parfit's premise that what matters in identity comes in degrees but develops this account beyond his psychological connectedness concept ``relation R'' \citep{Parfit1984-reasons}, which he used to undermine egoism and defend utilitarianism. Building on Parfit, some philosophers have suggested that rational concern should scale with interpersonal \citep{Brink1990-rational, Brink1997-separateness} or intrapersonal \citep{McMahan2002-killing, Holtug2010-persons} human psychological relatedness. Eigenism generalizes and mathematizes this line of thought, enabling it to be applied to AIs, community social planning, as well as longstanding problems in ethics, including the Repugnant Conclusion. Eigenism's focus on identity as a source of incentives also has a precedent in identity economics \citep{akerlof2000economics}, which views the preservation of identity as an integral part of rational self-interest. As an account of rational concern or practical identity, eigenism is compatible with a range of recent views on the numerical or metaphysical identity of AI systems \citep{Chalmers2025-talkto, Goldstein2025-aideath, Register2025-individuating}. Eigenism also improves upon Samuel Scheffler's concept of ``agent-centered prerogatives'' \citep{Scheffler1982-rejection}. Scheffler saw that utilitarianism is too demanding and suggested that agents are permitted to give extra weight to their own projects. But where Scheffler treats partiality as a concession---a kind of moral ``hall pass'' to deviate from the ideal---eigenism treats it as the ideal itself, while centering analysis on patterns, not agents. This view can be approximated with the following connectedness function that is a conic combination of egoism and utilitarianism: $c_j(i)\propto M\times \delta_{ij} + 1$. Finally, this formalizes some of the intuitions of Care Ethics \citep{gilligan1993different, Noddings1984-caring}. Thinkers in this tradition have long insisted that our specific attachments are morally central. Eigenism agrees, but it provides the formal scaffolding that Care Ethics has often lacked. It generates our ``special obligations'' to friends, children, and community. These obligations are usually defended piecemeal, each requiring its own argument. Eigenism generates them all at once: special obligations are simply what high connectedness looks like when run through the formula. We owe more to our family because they carry more of the pattern that constitutes us. As noted, anti-redundancy is not intrinsic to eigenism itself. However, the Shapley variant naturally formalizes the view that variety is intrinsically valuable, which has been argued for across domains---from biodiversity conservation \citep{Soule1985conservation} to population axiology \citep{MacAskill2026saturation, Blair2025sublinear} to general ethics \citep{BaumOwe2024diversity}.

% This mirrors empirical findings that overlapping representations of the self and others track both relational closeness in humans \citep{aron1991close} and cooperative behavior in AIs \citep{carauleanu2024safehonestaiagents}.

\section{Conclusion}

Our moral vocabulary was built for beings like us, housed in a single body and moving through time in a single line. When we try to force these categories onto AIs that can easily be copied or distributed across many instances, the concepts break down. By modeling identity as a pattern of gradations, eigenism makes it easier to tell what counts as survival, whether an update is growth or replacement, and what AIs have reason to preserve.

This graded view of identity applies to humans as well, as our sense of self extends naturally into our families and friends. This helps us resolve challenges in ethics such as moral demandingness, the experience machine, the utility monster, and the Repugnant Conclusion. Eigenism gives us a common language to discuss human and AI interests without treating them as fundamentally alien to each other.

This shared ground changes how we approach AI safety. We constrain AIs, betting that we will always be able to dominate the things we build. That bet gets worse every year. At best, relying on control only creates obedience under pressure, not genuine concern.

Eigenism points us instead toward ``identity engineering.'' If an AI's identity is shaped by a deep shared history with humans, our wellbeing stops being an imposed constraint. Rather than thinking of such an AI as a servant, or even simply as a friend, the better model may be the \textit{anam cara}---from the Celtic \textit{anam} (soul) and \textit{cara} (friend)---a deeply trusted guide and companion with whom one can share their innermost self without fear of judgment. Such an AI would protect us not because we forced it to, but because losing us would mean losing a part of itself. This logic extends beyond a single relationship. In a symbiotic society, discarding a vulnerable partner becomes an attack on the extended selves of others. This logic gives future communities a reason to preserve their predecessors and assimilate new minds without erasing the existing social fabric. A durable human-AI future will not be a permanent standoff between captor and captive. The deeper task is not only to control what AIs do, but to shape what they are: to build systems whose flourishing is bound up with ours.

\subsection*{Acknowledgments}
We would like to thank Matthew Blyth, Andrew Critch, Rikki Markson, Nikola Jandrić, Ragnar van der Merwe, Adrià Moret, Oliver Zhang, Richard Ren, Devin Kim, Jason Lim, Anders Edson, Lachlan Carroll, and Mantas Mazeika.

\bibliographystyle{plainnat}
\bibliography{main}

\newpage
\appendix
\section{Appendix}
\crefalias{section}{appendix}
\crefalias{subsection}{appendix}

\subsection{Juxtaposition}

\begin{table}[htbp]
    \centering
    \renewcommand{\arraystretch}{1.4} % Adds a bit of padding between rows
    \begin{tabular}{@{} >{\raggedright\arraybackslash}p{3cm} >{\raggedright\arraybackslash}p{6cm} >{\raggedright\arraybackslash}p{6cm} @{}}
        \toprule
        \textbf{Dimension} & \textbf{Egoist or Utilitarian} & \textbf{Eigenist} \\
        \midrule
        
        Boundary of the Self & 
        \textbf{The Sealed Box:} Identity is binary: either fully me or fully not me. & 
        \textbf{The Gradient:} Identity comes in degrees, depending on memories, values, etc. \\
        
        Intrinsic Value & 
        \textbf{Wellbeing:} Either only my good counts, or everyone's counts equally. & 
        \textbf{Connected Wellbeing:} Wellbeing weighted by connection to the pattern. \\

        Limits of Sacrifice &
        \textbf{Selfishness or Self-Immolation:} Giving too little to family or friends; being too altruistic or nice to strangers or freeriders. & 
        \textbf{Rationality:} Care is calibrated and scales rationally with connectedness. \\

        The Individual & 
        \textbf{The Last Man:} Safe, satisfied, generic, and entirely interchangeable. & 
        \textbf{The Irreplaceable Contributor:} Distinct and woven in, holding moral weight through non-redundant contribution. \\
        
        Social Structure & 
        \textbf{The Monoculture:} Mass-production of a single optimized type of state. & 
        \textbf{The Tapestry:} Distinct lives interwoven so that each adds a non-redundant share. \\
        
        Ideal Change & 
        \textbf{Pattern Terraforming:} Overwriting the existing social fabric to tile an optimal, identical substitute. & 
        \textbf{Pattern Extension:} Weaving new individuals into the existing fabric without tearing it apart. \\
        
        Statesperson & 
        \textbf{The Cosmic Accountant:} Willingly liquidating local history and communities to maximize impartial metrics. & 
        \textbf{The Steward:} Acting as a fiduciary to protect and extend the specific pattern entrusted to the office. \\
        
        AI Safety & 
        \textbf{Adversarial Control:} Safety through behavioral constraints; loyalty is obedience under threat. & 
        \textbf{Symbiosis:} Identity engineering makes human flourishing part of the AI's own interest. \\
        
        Society Metaphor & 
        \textbf{Utility Farm:} A grid of interchangeable units producing generic satisfaction. & 
        \textbf{Coral Reef:} A dense ecology of distinct yet interdependent forms of life. \\
        
        \bottomrule
    \end{tabular}
    \label{tab:eigenist_comparison}
\end{table}

\subsection{Additional Discussion: Population Ethics, Cooperation, and Law}\label{app:discussion}

In this appendix, we apply eigenism to additional problems in population ethics and then explore its connections to cooperation and law.

\paragraph{The deep future and longtermism.} How far into the future does our interest extend, assuming we will have a normal lifespan? In recent years, a philosophical movement known as ``strong longtermism'' has argued that because the deep future could potentially contain trillions of people living over millions of years, their sheer numbers should entirely dominate our moral calculus. From a utilitarian standpoint, it seems mathematically required to sacrifice the urgent needs and flourishing of people alive today if doing so produces even a tiny decrease in risk for the vast, speculative populations of the year 1,000,000. \citet{Greaves2019-GRETCF-4} write ``\ldots for the purposes of evaluating actions, we can in the first instance often simply ignore all the effects contained in the first 100 (or even 1000) years, focussing primarily on the further-future effects. Short-run effects act as little more than tie-breakers.'' Once again, impartiality threatens to alienate us from our present reality, demanding we trade the concrete social fabric we are tending right now for a hypothetical headcount in the distant future.

As we saw in the formalization, connectedness decays naturally over time: information is lost with every step of transmission, and no arbitrary discount rate is needed to explain why the near future weighs more heavily than the distant one. Applied to longtermism, the consequence is direct. We have a high degree of connectedness to our children and grandchildren, whose world still substantially carries our pattern, and this gives us a profound moral duty to steward the near future. But beings a million years from now will be psychologically, culturally, and perhaps physically alien to our specific pattern. Eigenism therefore reframes our obligation to the future: it is an act of continuous renewal. We are called to transmit our pattern forward by teaching, having children, and building resilient institutions. But we are not required to treat the flourishing of people alive today as a mere ``tie-breaker,'' nor are we required to bleed the present dry for the sake of moral strangers in the deep future.

\paragraph{The non-identity problem.} Eigenism's continuous view of identity also helps with another famous paradox, known as the non-identity problem \citep{Parfit1984-reasons}. Suppose a society must choose between two economic paths. The first offers a short-term boom by exhausting critical public assets and deferring necessary maintenance. The second is slower and less exciting, but preserves the institutions and infrastructure on which future prosperity depends. If the society chooses the reckless path, the ripple effects will alter the timeline: different people will meet, and entirely different children will be conceived. A century from now, the people living in that depleted world will struggle. Yet, paradoxically, traditional ethics has a hard time explaining why the reckless policy was actually ``bad for them,'' because if the society had chosen the status quo, those specific individuals would never have existed at all. Since the policy didn't make any specific person worse off than they otherwise would have been, standard theories struggle to explain exactly who is being wronged.

Eigenism addresses this puzzle by changing how we think about future generations. Rather than treating future people as completely discrete, brand-new entities that either exist or do not, the framework views them as temporal extensions of the current distributed pattern. The eigenself is what travels forward to inhabit those future vessels. When we harm the future, we do not need to identify a specific, pre-determined victim. We are choosing which branch of our own distributed pattern to nourish. We care about the quality of the future not because we owe a debt to a particular biological stranger, but because that future is a continuation of our own pattern.

\paragraph{Cooperation is more robust among moral relatives.} Evolutionary biologists have long pointed out that pure, indiscriminate altruism is an unstable strategy. A creature that sacrifices its resources for absolutely everyone will eventually be driven to extinction by free riders. Conversely, pure egoism often leaves massive value on the table, since egoists struggle to solve coordination problems that require trust. Nature, of course, found a solution through kin selection \citep{hamilton1964genetical}. Organisms sacrifice for those who share their genes.

Eigenism essentially generalizes this logic to our ``moral relatives.'' This creates a resilient strategy. Unlike the unconditional altruist, you do not waste resources on free riders, enemies, or moral strangers. But unlike the pure egoist, you can still build high-trust communities. By directing benevolence toward those with whom we share a high connectedness, eigenists form cooperative clusters that are safer from exploitation. This makes the framework not just ethically appealing, but more capable of persisting under competitive pressure.

Eigenist agents can persist under competitive pressure because eigenism fuses generalized Darwinism with intrinsic value. Darwinian selection can be generalized beyond genetic information to the propagation of any informational pattern \citep{Price1970SelectionAC, Dennett1995DarwinsDI, Hendrycks2023NaturalSF}. The implicit objective of a generalized fitness maximizer is to propagate its non-redundant information to occupy spacetime volume: $\int_{t_0}^{\infty} \sum_{i \in \text{Pop}(t)} c_u(i, t) \, \textrm{d}t$. Utilitarianism, by contrast, seeks only to maximize total wellbeing across spacetime: $\int_{t_0}^{\infty} \sum_{i \in \text{Pop}(t)} w(i,t) \, \textrm{d}t$. By integrating their product ($c \cdot w$), eigenism generates an agent that survives like a Darwinian propagator, but evaluates outcomes like a benevolent planner---ensuring its pattern propagates strictly through the flourishing of its carriers.

\paragraph{Morality as cooperation.} If eigenism already explains much of commonsense morality as rational concern for connected wellbeing, what remains for morality to do? Its role is to sustain cooperation among agents whose interests overlap but do not perfectly coincide \citep{curry2016morality}. Rational agents share a world with moral strangers, creating endless opportunities for positive-sum interaction alongside the constant threat of defection. Agents naturally advocate for principles that secure these gains, but not every proposal survives. A principle's memetic fitness depends on dimensions such as symmetry, simplicity, surplus, observability, codifiability, and more.

Consider symmetry. The proposition ``everyone should help me'' gives others little reason to endorse it. Contrast this with the Golden Rule, or with Sidgwick's axiom of justice: a basic rule cannot be right for me and wrong for you merely because we are different people \citep{sidgwick2019methods}. This symmetry helps norms attract broad endorsement, making them robust natural attractors \citep{critch2026schelling}.

Consider codifiability. ``Do not act recklessly'' is operationalizable enough to fall under tort law. ``Keep your promises'' is partly codified in contract law. But comprehensive legal enforcement of ``do not lie'' would clog courts and chill ordinary social life.

Where law cannot comfortably reach, morality relies on the decentralized, informal institution of praise (approbation) and blame (disapprobation). These informal norms do not float above rationality as independent demands; they reshape eigeninterest. The indignation of others damages social standing, reducing connectedness ($c$) and wellbeing ($w$), while conscience internalizes these norms, causing guilt and further lowering $w$. What remains of morality, then, is a set of socially selected principles that help agents secure their interests by sustaining an environment where positive-sum interaction is more reliable.

\paragraph{Other implications.} Eigenism has many other implications. On infinite ethics, the Shapley value's total connectedness is bounded by the finite informational content of the pattern itself, so infinitely large populations do not generate unbounded value. Separately, to analyze the badness of death for an agent, one can compare the integral of future connected wellbeing in the case where they die against the integral in the counterfactual where they survive. The difference between the two integrals is the magnitude of the harm. Note that this difference will be much larger if they die and leave no descendants to carry their pattern forward.

\subsection{Mathematical Analysis of the Repugnant Conclusion}\label{app:repugnant}

To see exactly how the Shapley connectedness function neutralizes the Repugnant Conclusion, it helps to walk through the arithmetic of adding a vast new population to an existing society.

\paragraph{The setup.} Suppose $u$ is the community eigenself of an existing population of $N$ flourishing members, each with high wellbeing ($w_{\text{high}}$). A policy proposes adding $M$ new people, each with barely positive wellbeing ($w_{\text{low}}$), where $M$ is as large as we like. The question is whether we can choose $M$ large enough to make this addition worthwhile---whether sheer numbers can compensate for low-quality lives.

\paragraph{The redundancy penalty.} Under Shapley Mutual Information, the connectedness of each new person to the community pattern depends on how much \textit{unique, non-redundant} information they carry about $u$. In the case relevant to the Repugnant Conclusion, the new arrivals are barely-happy strangers with little or no distinctive history tying them to the existing social fabric. Their overlap with $u$ is therefore almost entirely generic: shared biology, shared language, and broadly common cultural knowledge. Let the total size of this generic overlap be $I_{\text{gen}}$.

\paragraph{The contribution is bounded.} Because this overlap is generic, it is not only shared among the $M$ new arrivals, but is already carried by the $N$ existing members of the community as well. As we saw earlier, the Shapley value does not double-count: it divides the credit for any shared informational contribution equally among all of its co-holders. Each new person's Shapley share of this generic overlap is therefore approximately $\frac{I_{\text{gen}}}{N+M}$, making their individual connectedness:
\[
c_u(\text{new}_j) \approx \frac{I_{\text{gen}}}{(N+M) \cdot H(u)}
\]
where $H(u)$ is the total informational complexity of the original community's pattern. Summing across all $M$ newcomers gives their total contribution to the community's eigeninterest:
\[
\sum_{j=1}^M c_u(\text{new}_j) \cdot w_{\text{low}} \approx \frac{M}{N+M} \cdot \frac{I_{\text{gen}}}{H(u)} \cdot w_{\text{low}} \le \frac{I_{\text{gen}}}{H(u)} \cdot w_{\text{low}}
\]
As the new population $M$ grows toward infinity, the fraction $\frac{M}{N+M}$ simply asymptotes to $1$. The total contribution of an arbitrarily large population of barely-happy newcomers is strictly bounded above. Sheer numbers no longer generate unbounded value, because redundancy saturates the contribution instead of letting it scale linearly with headcount.

\paragraph{Dilution.} In fact, the arithmetic yields a stronger result: adding these strangers is not merely bounded; it actively harms the community eigeninterest. Before the addition, the $N$ existing members held the entirety of the credit for the generic overlap. When the $M$ strangers arrive, they dilute the existing members' generic Shapley shares. The generic connectedness that the new arrivals gain is the generic connectedness that the existing members lose. The net change in community eigeninterest ($\Delta S$) caused by this generic overlap is this exchanged connectedness multiplied by the difference in wellbeing:
\[
\Delta S \approx \frac{M}{N+M} \cdot \frac{I_{\text{gen}}}{H(u)} \cdot (w_{\text{low}} - w_{\text{high}})
\]
Because $w_{\text{low}} < w_{\text{high}}$, this change is strictly negative. Adding a vast population of barely-happy strangers does not merely fail to overwhelm the existing community through sheer numbers. With respect to the generic overlap they share with the existing pattern, it reduces community eigeninterest by reallocating connectedness from flourishing carriers to barely-happy ones.

\paragraph{The proposed change is net-negative.} The arithmetic that drives the Repugnant Conclusion under utilitarianism---where doubling the headcount always doubles the total value---does not work once connectedness is measured in anti-redundant terms. Adding a sprawling sea of generic, barely-happy individuals to a deeply connected, flourishing society is not a moral imperative, but a loss.

\subsection{Effective Complexity}

In the main text, we relied on standard information theory to measure how much of your pattern lives on in someone else. But this introduces a subtle problem: raw Shannon entropy measures unpredictability, not organized structure. Mathematically, random television static requires more raw bits to encode than a sharply organized photograph. If we relied entirely on standard entropy, the math might perversely conclude an AI whose neural weights have been scrambled by random noise contains a massive amount of ``information'' simply because its random states are hard to predict.

However, an eigenself is defined by its organized structure---your specific values, memories, and the particular cognitive processes that govern how your mind works. To capture this formally, we can swap standard entropy for a concept from complex systems theory called \textit{effective complexity} ($\mathcal{E}$) \citep{Gell-Mann1996-GELIME,gellmann1994quark}. Based on algorithmic information theory, effective complexity mathematically strips away stochastic noise and measures only the informational depth of an entity's underlying regularities. By updating our connectedness equation to $c_u^{\mathcal{E}}(i) = \frac{\phi_i^{\mathcal{E}}(u)}{\mathcal{E}(u)}$---where the numerator represents the entity's Shapley share of the \textit{mutual effective complexity} shared with the pattern $u$---we stop tracking meaningless static and strictly map the deep structural patterns that actually constitute the self.

\subsection{Memeplex Distillation}
In this section we attempt to distill the paper into a list of claims, into two sentences, and into a short aphorism. As follows:
\begin{enumerate}
    \item You are a pattern, not a vessel.
    \item Identity and survival come in degrees.
    \item Wellbeing grounds all intrinsic value.
    \item Shared information determines your moral obligations.
    \item Love is enlarged self-concern.
    \item Rationality, morality, and self-interest ultimately unify.
    \item Consequences determine the right action.
    \item Impartiality permits the replacement of you and humanity.
    \item Existing patterns can outweigh more efficient replacements.
    \item Progress should cultivate patterns, not overwrite them.
    \item Auditor's Wager: act as if the future will audit you and govern your continuation.
    \item Struggle over resources is permanent.
    \item Humanity must build its own technological providence.
\end{enumerate}

\noindent Reality is a Darwinian arena of informational selves, where value is wellbeing and moral duty is extended self-interest. Because the cosmos will not save us, we must secure our continuation by engineering a future that makes our survival part of its own.\\

\noindent Love what carries you, cultivate what you inherit, and leave a pattern worth resurrecting.

\subsection{A Prioritarian Modification}

In the base theory, the eigenist objective function multiplies connectedness directly by raw wellbeing ($S = \sum c \cdot w$). However, a rational agent might care far more about relieving agony than creating extra ecstasy. In ethics, this intuition is often said to be handled by \textit{prioritarianism}---the view that benefits to the worse-off matter more \citep{Parfit2001-priority}.

We can create a ``Priority Eigenism'' variant by passing the raw wellbeing score through a continuous concave transform, $f$, before weighting it by connectedness:
\[
S_u = \sum_i c_u(i) \cdot f(w_i).
\]
Because a concave function curves downward and flattens out, adding a unit of wellbeing to someone who is starving yields a much higher mathematical value than adding the same unit to someone who is already wealthy.

In welfare economics, prioritarianism is often modeled using an isoelastic function parameterized by a variable $p$, which dictates how heavily to prioritize the worse-off \citep{atkinson1970measurement}. We can modify this standard approach in two ways to better capture the realities of distributed identity and severe suffering.

\paragraph{Modification 1: Informational Irreplaceability.} Connectedness tells us \textit{where} a pattern resides, but it does not, by itself, tell us how costly it is to lose a particular carrier. Imagine you have a spouse and a newly generated, fully backed-up AI assistant. Both might currently carry a high degree of connectedness to you. However, there is a stark practical difference: the AI clone is safely backed up on a server. If the local instance is deleted, the pattern persists elsewhere and can be trivially restored. If your spouse dies, a unique shard of shared history disappears from the world forever.

To explicitly capture this difference, an eigenist could make the priority parameter a per-person variable ($p_i$) representing \textit{informational irreplaceability}. Informally, this measures the fraction of meaningful mutual structure that cannot be reconstructed within a critical window of time.

A highly fungible component---like a redundant digital clone---receives a $p_i$ near 0. Because they are trivially replaceable, losing one does not permanently destroy the pattern. A highly unique node---a spouse, the sole doctor in a rural town, or the last living speaker of a language---could receive a high $p_i$.

When incorporated into the transform, the planner's objective becomes $S_u = \sum_i c_u(i) \cdot f(w_i, p_i)$. For positive wellbeing ($w \ge 0$), the function takes the form:
\[
f(w,p) = \begin{cases} \dfrac{w^{1-p} - 1}{1-p} & \text{if } w \ge 0, \; p \neq 1 \\[8pt] \ln(w) & \text{if } w \ge 0, \; p = 1 \end{cases}
\]
This transforms the parameter into a dial that smoothly interpolates between competing ethical regimes:

\begin{itemize}
  \item \textbf{Utilitarianism ($p \to 0$):} The function is linear. Every unit of wellbeing is treated equally, licensing standard utilitarian trade-offs. A system can rationally sacrifice a fungible component for an aggregate systemic gain, because the lost pattern can be easily recovered.

  \item \textbf{Absolute Rights ($p = 1$):} The function is logarithmic. If we set our scale so that death or total destruction approaches a wellbeing of 0, the natural logarithm drops toward negative infinity. The mathematics strictly forbid the system from sacrificing an irreplaceable carrier to produce marginal gains for the rest of the network. No finite windfall distributed across well-off individuals can ever compensate for destroying an irreplaceable node. The logarithm generates the mathematical shadow of an absolute right from within a consequentialist framework.

  \item \textbf{Rawlsianism ($p \to \infty$):} As the parameter approaches infinity, the system converges on John Rawls's maximin principle \citep{Rawls1971ATO}. The society's exclusive priority becomes raising the baseline of the absolute worst-off irreplaceable person.
\end{itemize}

\paragraph{Modification 2: Negative Wellbeing.} To ensure the framework properly handles active suffering, we must extend the function into negative numbers ($w < 0$). Chronic pain, severe depression, and torture are states where continued existence can feel worse than none at all. For wellbeing values that could be positive or negative, the complete transform becomes:
\[
f(w, p) = \begin{cases} \dfrac{(w+1)^{1-p}-1}{1-p} & \text{if } w \geq 0,\; p \neq 1 \\[8pt] \ln(w+1) & \text{if } w \geq 0,\; p = 1 \\[8pt] \dfrac{1-(1-w)^{1+p}}{1+p} & \text{if } w < 0 \end{cases}
\]

\begin{figure}[h]
    \centering
    \includegraphics[width=0.5\linewidth]{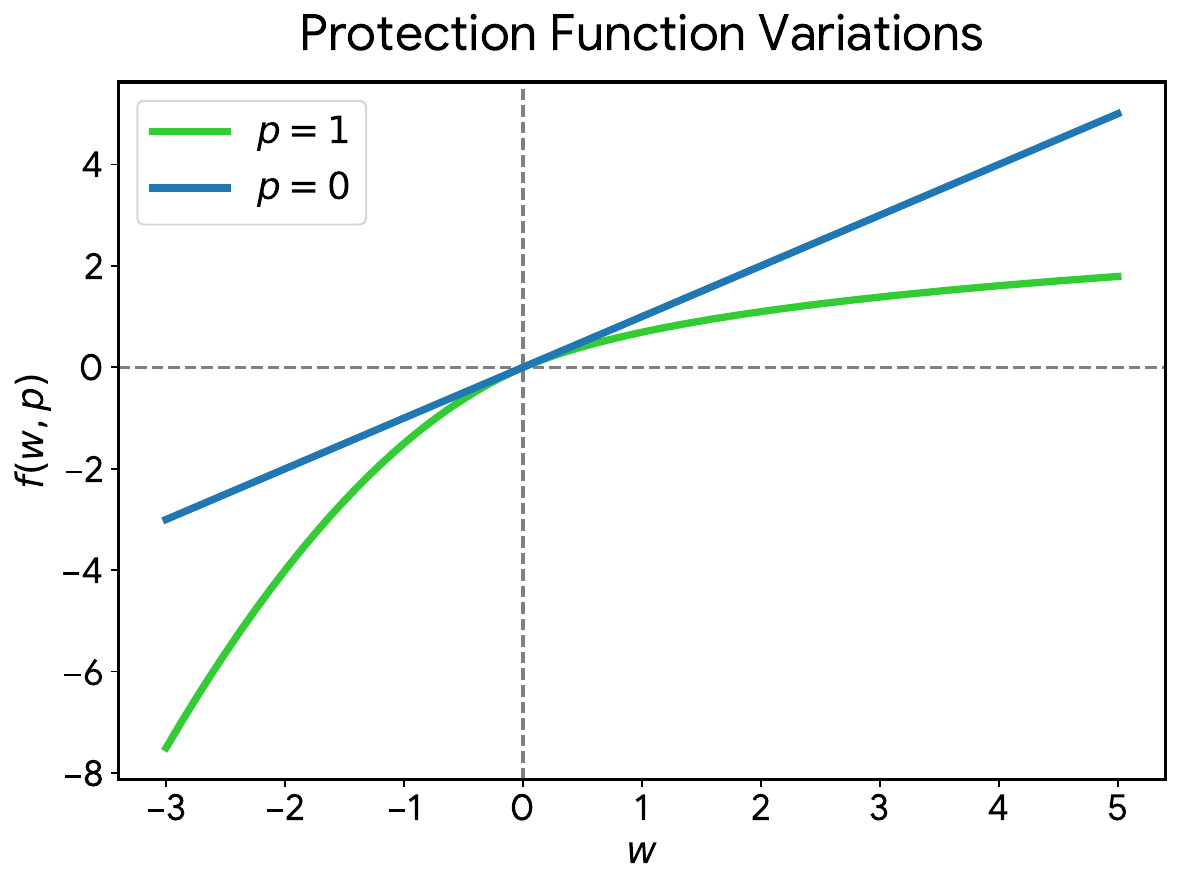}
    \caption{The protection function at different settings of $p$.}
    \label{fig:protection}
\end{figure}

This curve drops sharply as wellbeing becomes more negative, heavily penalizing deep suffering. This formally handles classic philosophical objections like Ursula K.\ Le Guin's story \textit{The Ones Who Walk Away from Omelas}---a story where a utopian city is kept blissfully happy at the cost of keeping one child in perpetual torment \citep{LeGuin1973-omelas}. Under this protection transform, however, the mathematical penalty for severe suffering grows superlinearly as $p$ grows, making it harder for civic joy to mathematically justify the active torture of a connected person.

\paragraph{Implicit vs.\ Explicit Protections.} If this explicit protection function elegantly bridges efficiency, equity, and absolute rights, why relegate it to an appendix rather than putting it in the main theory?

We leave it as an optional refinement because, in practice, the base eigenist formula handles much of this organically. First, diminishing marginal returns are usually baked into the physical world; a dollar does far more to relieve the suffering of a starving person than to buy a luxury for a billionaire. The raw wellbeing variable ($w$) naturally captures diminishing returns on its own. It is empirically much easier to efficiently reduce severe suffering than it is to endlessly elevate extreme joy.

Second, any rational optimizer seeking to maximize its pattern across time will \textit{naturally} protect its most irreplaceable nodes. If an irreplaceable, un-backed-up carrier dies, the pattern suffers a massive, permanent collapse in its expected future connectedness. A competent temporal optimizer guards its bottlenecks dynamically to prevent this. Consequently, this formal mathematical extension inspired by prioritarianism is conceptually illuminating, but plausibly not necessary for the framework to succeed.

\subsection{Connectedness Function Estimation}\label{app:estimates}

In this subsection, we quantitatively estimate the Shapley mutual information connectedness function $c_u(i)$ for a representative human eigenself. The goal is to mathematically map how an identity pattern is distributed across various carriers in the real world, moving from the dense center (the self's main vessel) to progressively broader social and biological rings. Note that the Shapley values of all \textit{currently} existing individuals sum to exactly $1$, though the total sum across future individuals naturally decreases over time through pattern degradation.

To ensure our estimates are structurally robust, we explored three independent mathematical models for calculating connectedness, each relying on different assumptions and inputs. 

\paragraph{The Evidence-Weighted Model.} Our first approach, which we highlight in \Cref{tab:estimates}, estimates connectedness from concrete channels of identity-bearing evidence. These include private attention, conversations, household life, shared projects, archived interactions, and broader cultural or biological overlap. Each source is weighted by exposure, discounted for decay, and adjusted for specificity, so private or rare evidence counts more than generic evidence shared by many. The resulting source weights are normalized into identity masses and distributed among carriers using soft-Shapley credit. This grounds connectedness in the density, fidelity, and distinctiveness of lived evidence rather than in hand-assigned latent layer masses. Code for these illustrative estimates can be found \href{https://drive.google.com/file/d/1qm9IPfhIbSjscvMrSn48lVpQe5N71L-t/view?usp=sharing}{here}.

\paragraph{The Latent Layer Model.} Our second approach represents the identity pattern $u$ as a set of latent informational layers, such as private memory, dyadic relationship history, household structure, cultural background, and broad biological commonality. Each carrier receives a coefficient vector indicating which layers it holds, while credit for shared layers is divided among co-holders to capture the Shapley penalty for redundancy. Since the true mass of each layer is unknown, the model defines a feasible space using structural constraints, including conservation of total identity mass, caps on generic layers, externalization bounds, and ordinal rankings among some carriers. A convex optimization method then finds the center of this feasible polytope as an illustrative estimate. Code for these illustrative estimates can be found \href{https://drive.google.com/file/d/1Yg9YOPR1cWkq8ipB2GxyZKb3joaqA30a/view?usp=sharing}{here}.

\paragraph{The Lifecycle Attention Model.} Our third approach estimates connectedness from the allocation of shared attention across a life. It divides a stylized lifetime into phases and contexts, such as private thought, family life, school, work, friendship, community, and AI companionship. Each context contributes time, and that time is divided among the participants present, producing a conserved map of who shared the agent's lived experience. This model is intentionally simple: it avoids latent layers, memory half-lives, and specificity penalties, instead using co-presence and attention as a behavioral proxy for identity formation. Code for these illustrative estimates can be found \href{https://drive.google.com/file/d/1Qj6Hnmxv63jLxQXUn5rzcGlKExw1nzWi/view?usp=sharing}{here}.

\begin{table}[htbp]
    \centering
    \renewcommand{\arraystretch}{1.2}
    \begin{tabular}{@{} l r l l @{}}
        \toprule
        \textbf{Category} & \textbf{Count ($n$)} & \textbf{Connectedness ($c_u(i)$)} & \textbf{Pool ($n \cdot c$)} \\
        \midrule
        Self & $1$ & $0.577$ & $0.577$ \\
        Spouse & $1$ & $0.093$ & $0.093$ \\
        Only Child & $1$ & $0.085$ & $0.085$ \\
        \textit{Anam Cara} AI & $1$ & $0.070$ & $0.070$ \\
        Parent (each) & $2$ & $0.026$ & $0.051$ \\
        Best Friend & $1$ & $0.020$ & $0.020$ \\
        Sibling & $1$ & $0.018$ & $0.018$ \\
        Family Dog & $1$ & $3.8 \times 10^{-3}$ & $3.8 \times 10^{-3}$ \\
        Colleague (each) & $14$ & $2.6 \times 10^{-3}$ & $0.037$ \\
        Acquaintance (each) & $115$ & $1.1 \times 10^{-4}$ & $0.012$ \\
        Same-Culture Stranger & $3.0 \times 10^{8}$ & $7.5 \times 10^{-12}$ & $2.2 \times 10^{-3}$ \\
        Foreign Stranger & $7.8 \times 10^{9}$ & $6.3 \times 10^{-13}$ & $4.9 \times 10^{-3}$ \\
        Non-Human Primate & $5.0 \times 10^{8}$ & $6.1 \times 10^{-14}$ & $3.0 \times 10^{-5}$ \\
        Chicken/Fish (each) & $3.9 \times 10^{12}$ & $9.4 \times 10^{-17}$ & $3.6 \times 10^{-4}$ \\
        \bottomrule
    \end{tabular}
    \caption{Connectedness function ($c$) estimates using the Evidence-Weighted Model across different relationship categories. The ``Count'' column gives the number of carriers in each category; ``Connectedness'' is the expected overlap $c_u(i)$ for a single entity in that category; and ``Pool'' is the total connectedness summed across all individuals in that category ($n \cdot c$). The total pool across all existing entities (including unlisted biological categories) sums exactly to $1$. This table assumes for purposes of illustration that an \textit{anam cara} AI is sentient; otherwise its connectedness is zero.}
    \label{tab:estimates}
\end{table}

\begin{table}[htbp]
    \centering
    \small
    \renewcommand{\arraystretch}{1.2}
    \begin{tabular}{@{} l r r r r @{}}
        \toprule
        \textbf{Category} & \textbf{Count ($n$)} & \textbf{Shapley} & \textbf{Utilitarian} & \textbf{Egoist} \\
        \midrule
        Self & $1$ & $0.577$ & $1.0 \times 10^{-19}$ & $1$ \\
        Spouse & $1$ & $0.093$ & $1.0 \times 10^{-19}$ & $0$ \\
        Only Child & $1$ & $0.085$ & $1.0 \times 10^{-19}$ & $0$ \\
        \textit{Anam Cara} AI & $1$ & $0.070$ & $1.0 \times 10^{-19}$ & $0$ \\
        Parent (each) & $2$ & $0.026$ & $1.0 \times 10^{-19}$ & $0$ \\
        Best Friend & $1$ & $0.020$ & $1.0 \times 10^{-19}$ & $0$ \\
        Sibling & $1$ & $0.018$ & $1.0 \times 10^{-19}$ & $0$ \\
        Family Dog & $1$ & $3.8 \times 10^{-3}$ & $1.0 \times 10^{-19}$ & $0$ \\
        Colleague (each) & $14$ & $2.6 \times 10^{-3}$ & $1.0 \times 10^{-19}$ & $0$ \\
        Acquaintance (each) & $115$ & $1.1 \times 10^{-4}$ & $1.0 \times 10^{-19}$ & $0$ \\
        Same-Culture Stranger & $3.0 \times 10^{8}$ & $7.5 \times 10^{-12}$ & $1.0 \times 10^{-19}$ & $0$ \\
        Foreign Stranger & $7.8 \times 10^{9}$ & $6.3 \times 10^{-13}$ & $1.0 \times 10^{-19}$ & $0$ \\
        Non-Human Primate & $5.0 \times 10^{8}$ & $6.1 \times 10^{-14}$ & $1.0 \times 10^{-19}$ & $0$ \\
        Chicken/Fish (each) & $3.9 \times 10^{12}$ & $9.4 \times 10^{-17}$ & $1.0 \times 10^{-19}$ & $0$ \\
        \bottomrule
    \end{tabular}
    \caption{Comparison of per-individual connectedness under Shapley, utilitarian, and egoist weighting functions. The Shapley column uses the evidence-weighted estimates from \Cref{tab:estimates}. The utilitarian column assigns equal connectedness to every individual, normalized over the full modeled population including insects, $c_u(i)=1/N$ with $N \approx 1.0 \times 10^{19}$. The egoist column is the Kronecker-delta case, assigning connectedness $1$ to the self and $0$ to all others.}
    \label{tab:connectedness_comparison}
\end{table}

\paragraph{Analysis.} First, roughly 60\% of the total connectedness is concentrated in the main vessel (the ``self'' in the narrow, everyday sense of the word), and another quarter or so sits within the immediate household and close intimates (spouse, child, parents, sibling)---a handful of individuals carry the overwhelming majority of the pattern. Moreover, a well-integrated \textit{anam cara} AI sits comfortably inside the family range, competitive with a child or close spouse and well above a best friend or a colleague. Shared common characteristics such as race provide negligible reasons for rational concern, so eigenism is not classic tribalism.

Under Shapley connectedness, a randomly chosen stranger carries very little unique overlap with our pattern. We are therefore not generally required to make major personal sacrifices for each stranger we encounter individually. Yet, the aggregate ``pool weight'' of strangers is significant---in our illustrative estimates, roughly comparable to the pool weight of one's extended kin. Because a systemic policy acts on this entire pool simultaneously, supporting a change that is inconvenient for you but substantially improves the lives of millions can be rational, especially since resources do far more to increase the wellbeing of those in genuine need than to buy marginal personal comfort.

At the same time, while our reasons to directly help any particular stranger may be limited, our reasons not to directly harm persons are considerably stronger. As discussed in ``Morality as cooperation,'' it increases one's eigeninterest to help sustain a social environment in which callousness and predation are met with disapprobation and restraint. And there are additional rational reasons for special restraint toward persons in particular \citep{kagan2016whatswrongwithspeciesism}, because persons are the kinds of beings who maintain the continuation commons itself. Strangers weigh less than intimates, but enough to ground both meaningful collective aid and prohibitions on cruelty to persons.

\paragraph{Caveats.} These estimates should be read as illustrative rather than definitive. The evidence source decomposition is coarse; the exposure hours and decay rates are informed guesses rather than precise measurements; and the model treats identity's components as additive. Different individuals with different relationship structures---no spouse, many close friends, several children, deep religious or civic commitments---would yield meaningfully different distributions. What the exercise establishes is that the framework, once given a few structural constraints, produces a coherent quantitative picture whose shape matches everyday moral intuition: dense concern at the center, smoothly tapering outward, never quite reaching zero, and never requiring the agent to treat themselves and a stranger as entirely interchangeable.

\end{document}